# Anomalous Structural Response of Quasi-One-Dimensional Antiferromagnetic Metal $KMn_6Bi_5$ under high pressure

Hanming Ma [1,2=], Qingxin Dong [1,2=], Tong Shi [1,3], Xiaoli Ma [1,2], Zhongjin Wu [1,2], Shaoheng Ruan [1,2], Pengtao Yang [1,2], Zhaoming Tian [3], Jianping Sun [1,2], Yoshiya Uwatoko [4,5], Genfu Chen [1,2], Xiaohui Yu [1,2], Bosen Wang [1,2]*, and Jinguang Cheng [1,2]*

[1] Beijing National Laboratory for Condensed Matter Physics and Institute of Physics, Chinese Academy of Sciences, Beijing 100190, China

[2] School of Physical Sciences, University of Chinese Academy of Sciences, Beijing 100049, China

[3] Wuhan National High Magnetic Field Center and School of Physics, Huazhong University of Science and Technology, Wuhan 430074, China

[4] Department of Advanced Materials Science, Graduate School of Frontier Sciences, University of Tokyo, Kashiwa, Chiba 277-8581, Japan

[5] Department of Natural Sciences, Faculty of Science and Engineering, Tokyo City University, Setagaya-ku, Tokyo 158-8557, Japan

*Corresponding authors: bswang@iphy.ac.cn (BSW); jgcheng@iphy.ac.cn (JGC)

## Abstract

We report high-pressure single-crystal X-ray diffraction measurements on the quasi-one-dimensional (Q1D) antiferromagnetic metal $KMn_6Bi_5$ up to 12.5 GPa, revealing the detailed pressure evolution of its atomic coordination environment. We find that the lattice exhibits pronounced anisotropic compressibility-the relative changes in the *a* and *b* lattice parameters reach $a/a_0 \approx 0.91$ and $b/b_0 \approx 0.94$ at 12.5 GPa-and a distinct structural anomaly emerges near 11 GPa without any symmetry-breaking. Detailed structural analysis further uncovers an anomalous hardening of the Mn nanotubes between 5 and 11 GPa, followed by a configuration optimization of the Mn/Bi nanotubes around 11 GPa. These features correlate closely with the reported pressure-temperature phase diagram of $KMn_6Bi_5$ and compare favorably with the chemical pressure effects induced by substituting K with Na, Rb, or Cs. Our findings provide key microscopic insights into how coordination environment modulation governs the stability of electronic orders in low-dimensional systems.

## Introduction

Quantum criticality occurs when a long-range antiferromagnetically ordered state is suppressed continuously to zero temperature by a non-thermal parameter, often giving rise to superconductivity mediated by enhanced magnetic fluctuations near the quantum critical point (QCP) [1-13]. In many cases, structural details near the QCP remain obscure beyond a smooth volume change and an apparently unchanged crystal symmetry. Yet, the suppression of long-range antiferromagnetic (AFM) order typically involves modifications of exchange interactions through changes in interatomic distances and bond angles [14-18]. This is particularly relevant for two-dimensional and quasi-one-dimensional (Q1D) magnets, where inter-layer or inter-chain couplings can evolve in complex and often anisotropic ways. In this regard, single-crystal X-ray diffraction (SXRD) thus becomes a key tool for extracting structural characteristics near QCPs, providing essential insights into the driving forces of phase transitions and their microscopic mechanisms, and offering valuable guidance for the design of new quantum materials.

In the present study, we focus on a Q1D AFM metal $KMn_6Bi_5$ that was recently found to show unconventional superconductivity near the QCP and represents the first case of ternary Mn-based superconductor [12, 19-22]. As shown in Fig. 1(a), it crystallizes in a monoclinic structure (space group $C2/m$) featuring infinite $[Mn_6Bi_5]^-$ nanotubes extending along the crystallographic $b$ direction. Each nanotube consists of an outer Bi tube, an inner Mn tube, and a central Mn chain; both the Bi and Mn tubes are constructed by stacking Mn and Bi pentagons within the $ac$ plane. Previous studies have suggested that Mn atoms at the pentagon sites and those along the central chain exhibit pronounced charge disproportionation, with marked different localized magnetic moments of ~ 2.46 $\mu_B$ and ~ 0.29 $\mu_B$, respectively [22].

At ambient pressure (AP), $KMn_6Bi_5$ undergoes an AFM transition with a Néel temperature $T_N \approx 75$ K. Neutron scattering experiments indicate that the magnetic order likely originates from a combination of multiple exchange mechanisms, including Mn-Mn direct exchange, Mn-Bi-Mn super-exchange, and interactions between conduction electrons and localized Mn-3$d$ electrons due to complex structural configurations [13, 22]. Notably, low-temperature SXRD measurements reveal clear magnetoelastic effects: the $a$ and $c$ axes expand, the $b$ axis contracts, and the monoclinic angle $\beta$ increases upon cooling through $T_N$. These observations point to a strong spin-lattice coupling [22], and suggest that the lattice parameters carry vital information for understanding its magnetic state. Under external pressures, $KMn_6Bi_5$ exhibits a rich phase diagram, as summarized in Fig. 1(b). $T_N$ initially rises to ~83 K at 4 GPa, then drops markedly to ~ 60 K at 6 GPa, gradually decreases to ~ 50 K at 11 GPa, and finally collapses near 13 GPa, where superconductivity emerges with an optimal onset temperature of $T_c \approx 9.3$ K at 14 GPa [13]. First-principles calculations on the related compound $RbMn_6Bi_5$ and $CsMn_6Bi_5$ indicate that Mn-3*d* states dominate the density of states near the Fermi level, highlighting the key role of the Mn coordination environment in the electronic properties of this family [12, 21].

Although the pressure evolution of AFM order and superconducting phase in $KMn_6Bi_5$ has been established, a clear microscopic structural perspective that traces the evolution of electronic orders under pressure is still lacking. In particular, a detailed knowledge of interatomic distances is essential for characterizing the local coordination environments of Mn atoms and for tracking their evolution under compression. Such information is expected to be instrumental in elucidating the exchange interactions responsible for pressure-induced magnetic collapse. In this work, we have performed high-pressure SXRD measurements on $KMn_6Bi_5$ to investigate its structural evolution at the atomic scale. By tracking the pressure dependence of lattice parameters and interatomic distances, we clarify how the local coordination evolves under compression and how these structural modifications ultimately influence the magnetic properties in this Q1D system.

## Experimental

High-quality single crystals of $KMn_6Bi_5$ were synthesized following the procedures reported in [10, 12, 13]. SXRD measurements were performed at 300 K on a Rigaku XtaLAB Synergy-R diffractometer equipped with a HyPix detector and Mo Kα radiation ($\lambda$ = 0.71073 Å). The structure was solved using Olex2 with the SHELXT (2018/2) structure solution algorithm and refined by full-matrix least-squares methods on $F^2$ using SHELXL (2018/3) [23, 24]. High-pressure SXRD experiments were carried out with a diamond anvil cell (DAC) having a culet diameter of ~ 300 μm. A rhenium gasket was pre-indented from an initial thickness of 250 μm to ~70 μm, and a hole of 200 μm diameter was drilled in the center to serve as the sample chamber. A $KMn_6Bi_5$ single crystal with typical dimensions of ~ 180 μm× 60 μm× 40 μm was loaded into the DAC together with a small ruby sphere for pressure calibration. A 4:1 methanol-ethanol mixture was employed as the pressure-transmitting medium to ensure quasi-hydrostatic conditions. The pressure was determined using the ruby fluorescence method [25].

## Results

### Crystal structure and lattice parameters under pressure

SXRD experiments were first performed on $KMn_6Bi_5$ at AP outside of DAC; the corresponding diffraction pattern is displayed in Fig. 2(a). The sharp, well-defined diffraction spots confirm the high quality of the crystals used in this study. At AP, the diffraction data were successfully refined in the monoclinic space group $C2/m$, yielding unit-cell parameters $a$ = 22.9934(10)Å, $b$ = 4.6147(1)Å, $c$ = 13.3944(6)Å, $\alpha$ = $\gamma$ = 90°, $\beta$ = 124.591(6)°, and a unit-cell volume $V$ = 1170.01(10)Å$^3$ (see Table 1). The refined structure and lattice parameters are in good agreement with previous report [19]. After confirming the sample quality, we then carried out high-pressure SXRD measurements using a DAC up to ~12.5 GPa. The diffraction patterns collected at 0 and 12.53 GPa in DAC are presented in Figs. 2(b)-(c) as representative examples, demonstrating that the high sample quality is maintained under compression. These diffraction patterns can also be described with the same *C*2/*m* space group without showing structural phase transition in the studied pressure range.

The pressure dependence of the structural parameters is presented in Figs. 2(d-h). As can be seen, the lattice parameters *a*, *b*, and *c* decrease smoothly and monotonically with increasing pressure up to ~11 GPa, above which a distinct structural anomaly emerges: both *a* and *b* axes undergo an abrupt contraction, while the monoclinic angle $\beta$ exhibits a discontinuous increase. Notably, the *c* axis shows no discernible anomaly over the entire pressure range. These changes together lead to a negligible change in the unit-cell volume *V* at ~ 11 GPa, Fig. 2(g). Fitting the $V(P)$ data to the third-order Birch-Murnaghan equation of state [26] yields a small bulk modulus $B_0$ = 19.7(5) GPa as shown in Fig. 2(h).

The relative changes $a/a_0$, $b/b_0$, $c/c_0$ are plotted as functions of pressure in Figs. 2(d-f). The *a* axis is the most compressible, reaching $a/a_0 \approx 0.91$ at 12.5 GPa, indicating that the inter-tube distance along the [100] direction is exceptionally sensitive to external pressure. In contract, the *b* axis (the nanotube direction) is considerably more rigid, exhibiting only $b/b_0 \approx 0.94$ at the same pressure. This pronounced axial compressibility anisotropy reflects the intrinsically anisotropic response of the $[Mn_6Bi_5]^-$ framework. Above 11 GPa, the abrupt contraction of the *a* and *b* axes, together with the discontinuous jump in the $\beta$, signals a significant structural modification. Tracing the evolution of the lattice parameters seen in Table 1, we infer that this rearrangement likely involves not only the radial compression of the Mn/Bi nanotubes but also inter-tube sliding or rotational degrees of freedom as shown below. Such changes represent a critical reconfiguration of the coordination environments within the Q1D framework, serving as a structural precursor to the eventual collapse of the magnetic order in $KMn_6Bi_5$.

**The evolution of bond lengths and angles under pressure**

To gain deeper insight into the pressure-induced structural evolution, we examined the pressure dependence of the bond lengths and angles within the $KMn_6Bi_5$ framework, including the nanotube geometry and key interatomic distances. To quantitatively describe the structural modifications and changes in the local coordination environment, we define the following parameters, as illustrated in Figs. 3(a) and 4(a): (1) the radii of the Mn and Bi nanotubes, $r_{\mathrm{Mn}}$ and $r_{\mathrm{Bi}}$; (2) the orientation angles of the Mn and Bi nanotubes with respect to the *a* axis, $\Delta\theta_{\mathrm{Mn}}$ and $\Delta\theta_{\mathrm{Bi}}$; since the angle between the crystallographic plane defined by the Mn *i* ($i$ = 1 to 5) atoms and the *a* axis satisfies the relationship $\theta_{\mathrm{Mn}\,i} = \Delta\theta_{\mathrm{Mn}\,i} + i \times (360°/5)$ ( $i$ = 1 to 5), we denote the average offset angle, $\Delta\theta_{\mathrm{Mn}} = \langle\Delta\theta_{\mathrm{Mn}\,i}\rangle$, as the orientation angle of the Mn nanotubes. The same definition applies to the Bi nanotube orientation angle, $\Delta\theta_{Bi}$; (3) the distance between the central Mn6 atoms, $d_{Mn\text{-}Mn\ I}$; (4) the distance between neighboring Mn atoms at adjacent pentagon sites, $d_{Mn\text{-}Mn\ II}$; (5) the distance between the central and pentagon-site Mn atoms, $d_{Mn\text{-}Mn\ III}$; (6) the distance between neighboring Mn atoms within the same pentagon site, $d_{Mn\text{-}Mn\ IV}$; and (7) the inter-tube Bi-Bi distances, $d_{Bi\ I}$ and $d_{Bi\ II}$. Due to the high positional degrees of freedom within the *C2/m* space group, the specific interatomic distances associated with the pentagon sites for both Mn and Bi are not exactly the same. Thus, the $d_{Mn\text{-}Mn\ II}$, $d_{Mn\text{-}Mn\ III}$, $d_{Mn\text{-}Mn\ IV}$, $d_{Bi\ I}$ and $d_{Bi\ II}$ discussed here represent the averaged values. The entire set of individual bond lengths are summarized in Supplemental Material [27].

We begin with the geometric evolution of the $[Mn_6Bi_5]^-$ nanotube framework. A striking feature is the divergent pressure response of the inner and outer nanotubes. As shown in Fig. 3(b), $r_{Bi}$ undergoes a smooth, monotonic contraction up to 11 GPa. In contrast, $r_{Mn}$ exhibits a distinct kink near 5 GPa, followed by an anomalous hardening, a plateau-like regime in which $r_{Mn}$ remains nearly constant at ~ 2.30 Å up to 11 GPa. Such a stiffening of the inner Mn tube is effectively masked by the continuous compression of the outer Bi shell, rendering the anomaly virtually imperceptible in the macroscopic unit-cell parameters. Consequently, the radial separation ($r_{Bi}$-$r_{Mn}$) reaches its minimum around 11 GPa, indicating a significant increase in internal strain. Intriguingly, above 11 GPa, both $r_{Mn}$ and $r_{Bi}$ undergo rapid synchronous contraction, resulting in an anomalous enlargement of ($r_{Bi}$-$r_{Mn}$), which should modify the monotonic evolutions of exchange interactions via the Mn-Bi-Mn pathways.

The most striking finding of the present work is the sudden rotation of both Mn and Bi nanotubes at ~11 GPa as shown in Fig. 3(c). From AP to 11 GPa, both $\Delta\theta_{Mn}$ and $\Delta\theta_{Bi}$ remain at a value of approximately +13°, and then sharply decrease to nearly -13° above 11 GPa. This indicates a significant counterclockwise rotation of about 26° for both Mn and Bi nanotubes. Consequently, the inter-tube super-exchange pathways have undoubtedly undergone significant modifications. The collective high-pressure SXRD results point to an exotic hardening behavior of the Mn nanotubes and reveal that the isostructural phase transition in $KMn_6Bi_5$ involves a pronounced orientational modification of the Mn and Bi nanotubes. These findings suggest that both the intra-tube Mn-Mn and inter-tube Mn-Bi interactions are strongly modified.

We further analyzed the pressure dependence of the intra-tube Mn-Mn and inter-tube Bi-Bi interatomic distances, which govern the direct Mn-Mn exchange and the indirect Mn-Bi-Mn exchange interactions. As shown in Fig. 4(b), the Mn-Mn interatomic distances exhibit a highly anisotropic response. Whereas the axial distances ($d_{Mn\text{-}Mn\ I}$ and $d_{Mn\text{-}Mn\ II}$) decrease monotonically, the circumferential distances ($d_{Mn\text{-}Mn\ III}$ and $d_{Mn\text{-}Mn\ IV}$) show a pronounced slope change near 5 GPa and remain remarkably pressure insensitive up to 11 GPa. The hardening of the $d_{\text{Mn-Mn III}}$ and $d_{\text{Mn-Mn IV}}$ directly accounts for the plateau-like regime in the $r_{\text{Mn}}$ [Fig.3(b)] and indicates that the exchange interactions among the pentagon-site Mn atoms within the *ac* plane become exceptionally robust over the 5-11 GPa range. This behavior likely stems from the higher 3*d* electron occupancy on the pentagon-site Mn, which enhances electronic correlations and leads to the observed highly anisotropic exchange rigidity. Upon further compression above 11 GPa, all Mn-Mn distances undergo a sharp contraction. Simultaneously, the Bi-Bi distances $d_{\text{Bi I}}$ and $d_{\text{Bi II}}$, which decrease smoothly up to 11 GPa, begin to diverge [Fig.3(d)]: while $d_{\text{Bi I}}$ remains essentially constant at ~ 3.1 Å, $d_{\text{Bi II}}$ drops quickly, causing a reorientation of the nearest-neighbor inter-tube Bi-Bi direction. These observations provide clear evidence for pressure-induced modification of the magnetic exchange interactions, establishing a vital microscopic link to the collapse of AFM order and to the emergence of SC at magnetic QCP.

## Discussions

The above results provide a coherent picture of pressure-driven evolution of the AFM order in $KMn_6Bi_5$. In the low-pressure regime (0-5 GPa), the uniform compression of the Mn coordination environment enhances the 3*d* exchange interactions, giving rise to the initial increase of $T_N$. Upon entering the intermediate regime (5-11 GPa), the behavior becomes markedly nonmonotonic. Given the magnetoelastic effects reported near $T_N$, wherein the *a* and *c* axes expand while the *b* axis contracts [22], the anomalous hardening of the circumferential Mn-Mn bonds likely suppresses this necessary magnetoelastic coupling, thereby driving the abrupt drop in the $T_N$ above 5 GPa. In this regime, while the direct Mn-Mn exchange remains relatively stable owing to the bond hardening, the continuously evolving Mn-Bi-Mn super-exchange interactions probably underlie the gradual decrease of $T_N$ upon further compression. Above 11 GPa, both the circumferential and axial Mn-Mn bonds soften abruptly-a concomitant collapse of the inner Mn nanotube that is most plausibly induced by a significant hybridization of the 3*d* electrons at the pentagon-site Mn atoms. Under this scenario, the enhanced 3*d* hybridization can account for the collapse of AFM order, and the concomitant increase in metallic character at those Mn sites facilitates the emergence of superconducting state. This scenario, in which the suppression of magnetic order and the enhanced 3*d* itinerancy act as precursors to superconductivity, has been observed in other 3*d*-electron systems, including the helical AFM metals MnP and CrAs [4, 5].

Based on these findings, we highlight several points that deepen our understanding. First, we identify a pronounced structural modification-namely, the hardening of the Mn nanotube-that remains concealed within the macroscopic lattice parameters. The marked unit-cell anomalies observed near 11 GPa are driven by the structural response of both the Mn and Bi nanotubes. Notably, $r_{\text{Mn}}$ and $r_{\text{Bi}}$ display strikingly different pressure dependencies between 5 and 11 GPa: the Mn nanotube exhibits a clear hardening effect, whereas the outer Bi nanotube undergoes smooth, continuous compression up to 11 GPa. The simultaneous contraction of $r_{\mathrm{Mn}}$ and $r_{\mathrm{Bi}}$, together with the abrupt jumps in $\Delta\theta_{\mathrm{Mn}}$ and $\Delta\theta_{\mathrm{Bi}}$, indicates that the rearrangement involves a complex combination of nanotube contraction and relative rotation between adjacent tubes, rather than a simple positional translation. This behavior reflects a highly anisotropic evolution of the Mn-Mn exchange interactions within the Q1D structure, providing a critical bridge between the Mn coordination environment and the underlying 3*d*-electron ordering.

Second, the structural responses uncovered here can be broadly extended to the isostructural $A\mathrm{Mn_6Bi_5}$ family ($A$ = Na, K, Rb, and Cs) [13, 20, 21, 28]. Alkali-metal substitution introduces varying degrees of chemical pressure, which manifest as macroscopic changes in the lattice parameters. However, inspection of the local nanotube geometry reveals that for $A$ = K, Rb, and Cs, the radial dimensions of the inner Mn nanotubes undergo only negligible modifications [12, 21]. This observation rationalizes the similar pressure-temperature phase diagrams exhibited by these isostructural compounds [13, 21, 28]: the outer Bi nanotube effectively shields the inner Mn framework from the structural perturbations induced by chemical pressure, thereby preserving a nearly identical local coordination environment for the Mn atoms. The situation is distinctly different for $\mathrm{NaMn_6Bi_5}$. The substitution of the considerably smaller $\mathrm{Na^+}$ ion introduces a pronounced pre-compression effect that exceeds the shielding capacity of the Bi nanotube [29]. This structural vulnerability exposes the inner Mn framework to significant modifications, explaining why the magnetic and electronic states of $\mathrm{NaMn_6Bi_5}$ are exceptionally sensitive to external pressure [20]. Given that theoretical calculations identify 3*d* orbital modifications as the primary driving force for the AFM spin-flip and the emergence of superconductivity [12, 21], our work provides a comprehensive structural basis for understanding these electronic evolutions. These insights illustrate the complex interplay between local structural coordination and electronic phase stability in a wide range of low-dimensional correlated systems.

## Conclusion

In summary, we have established a comprehensive structural basis for pressure-induced magnetic and electronic evolution in $\mathrm{KMn_6Bi_5}$ by performing high-pressure single-crystal X-ray diffraction measurements up to 12.5 GPa. The lattice parameters exhibit pronounced anisotropic compressibility, with a distinct structural anomaly emerging near 11 GPa in the absence of any space-group symmetry change. A detailed analysis of bond distances and angles reveals that the underlying structural response is fundamentally governed by pressure-driven reconfiguration of the concentric Mn and Bi nanotubes. In particular, two sets of Mn-Mn bonds display markedly different compressibilities, resulting in an anomalous hardening of the inner Mn nanotube radius between 5 and 11 GPa within the *ac* plane. Our results firmly establish that the pressure-induced modification of the Mn coordination environment serves as the decisive tuning parameter for the 3*d*-electron exchange interactions and the concomitant evolution of AFM order. Collectively, these findings provide essential microscopic structural insights into the electronic and magnetic phase diagram of $KMn_6Bi_5$ and highlight the intimate connection between the local coordination environment and the stability of electronic ground states in low-dimensional correlated systems.

## Data availability

All data are available from the corresponding author upon reasonable request.

## Acknowledgements

This work was supported by the National Key Research and Development Program of China (Grants No.2024YFA1408400, 2023YFA1607402, 2023YFA1406100, 2021YFA1400200, 2021YFA1401800), the National Natural Science Foundation of China (Grant No. 12025408, 12074414), the CAS President's International Fellowship Initiative (Grant No. 2024PG0003), the Outstanding Member of Youth Promotion Association of CAS (Grant No. Y2022004), the CAS Superconducting Research Project (SCZX-0101), and the International Young Scientist Fellowship of the Institute of Physics, CAS (Grant No. 202604). This work was carried out at the cubic anvil cell (CAC) station and High-pressure synergetic measurement station of the Synergetic Extreme Condition User Facility (SECUF, https://cstr.cn/31123.02.SECUF).

## Competing interests

The authors declare no competing interests.

## Figure captions

**Figure 1 Crystal structure and pressure-temperature phase diagram of $KMn_6Bi_5$.** (a) The schematic crystal structure of $KMn_6Bi_5$. Green, grey, purple, and pink spheres represent K, Bi, pentagon-site Mn, and central-chain Mn atoms, respectively. Some atoms and bonds are omitted for clarity. The black dashed line outlines the unit cell. (b) The pressure-temperature phase diagram of $KMn_6Bi_5$ adopted from Ref. [18]. The AFM and SC represent the antiferromagnetic order and superconductivity, respectively.

**Figure 2 Single-crystal XRD patterns and lattice compression of $KMn_6Bi_5$.** (a-c) The SXRD patterns at AP and high pressures; (d-h) Pressure dependence of the lattice

parameters $a$, $b$, $c$, $\beta$ and the unit-cell volume $V$. The dashed line in (h) represents the fitting results of Birch-Murnaghan equation of state, while the other dashed lines are guide to the eye.

**Figure 3 Pressure-induced structural anomaly in the nanotube framework.** (a) Definition of the structural descriptors: the nanotube radii ($r_{\mathrm{Mn}}$ and $r_{\mathrm{Bi}}$), orientation angles ($\Delta\theta_{\mathrm{Mn}}$ and $\Delta\theta_{\mathrm{Bi}}$), and inter-tube Bi-Bi distances ($d_{\mathrm{Bi\ I}}$ and $d_{\mathrm{Bi\ II}}$). Pressure evolution of (b) the nanotube radii, (c) orientation angles, and (d) the inter-tube Bi-Bi distances ($d_{\mathrm{Bi\ I}}$ and $d_{\mathrm{Bi\ II}}$), highlighting the anomalous plateau between 5 and 11 GPa and the orientation angles jump near 11 GPa.

**Figure 4 Pressure-induced structural anomaly in the Mn-Mn distances.** (a) Definition of the structural descriptors: intra-tube Mn-Mn distances ($d_{\mathrm{Mn\text{-}Mn\ I}}$, $d_{\mathrm{Mn\text{-}Mn\ II}}$, $d_{\mathrm{Mn\text{-}Mn\ III}}$, and $d_{\mathrm{Mn\text{-}Mn\ IV}}$); (b) Evolution of the intra-tube Mn-Mn distances as a function of pressure, revealing the highly anisotropic compressibility between the Mn-Mn bonds.

# Figure 1

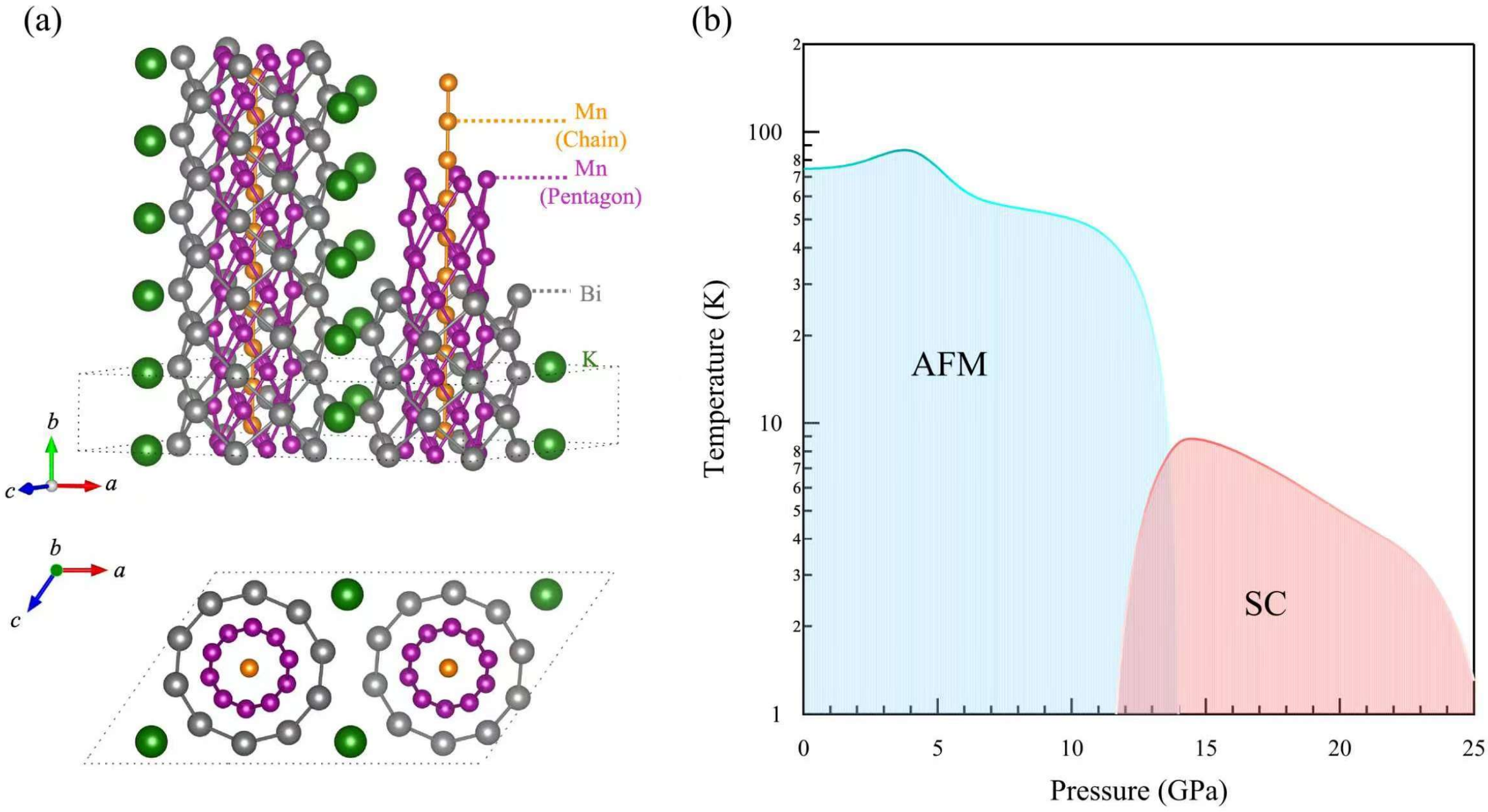

(a)
Mn
(Chain)
Mn
(Pentagon)
Bi
K
b
c
a
b
a
c
(b)
Temperature (K)
Pressure (GPa)
AFM
SC
100
10
1
0
5
10
15
20
25

# Figure 2

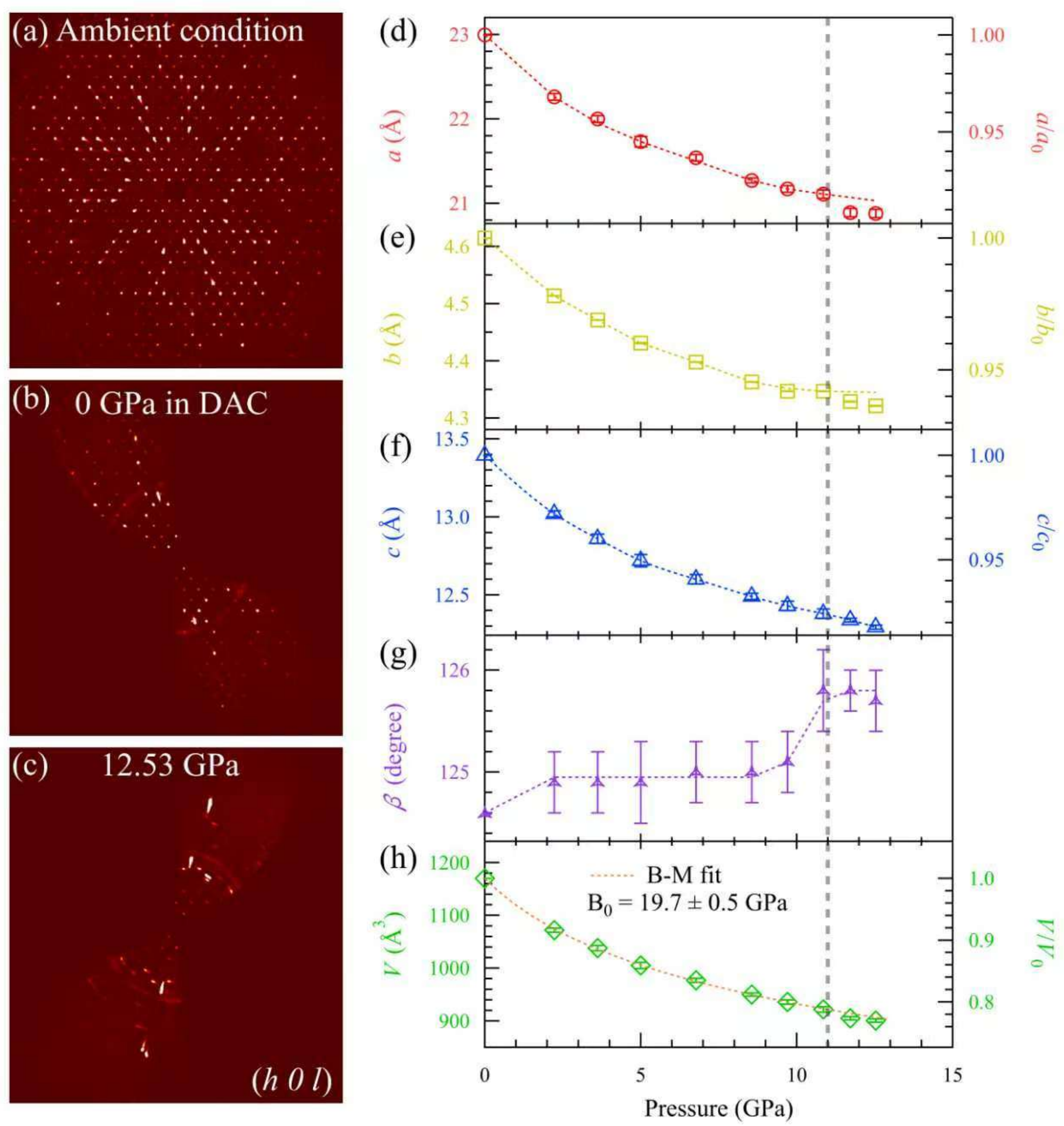

(a) Ambient condition
(b) 0 GPa in DAC
(c) 12.53 GPa
(h 0 l)
(d)
(e)
(f)
(g)
(h)
a (Å)
a/a0
b (Å)
b/b0
c (Å)
c/c0
β (degree)
V (Å3)
V/V0
B-M fit
B0 = 19.7 ± 0.5 GPa
Pressure (GPa)

# Figure 3

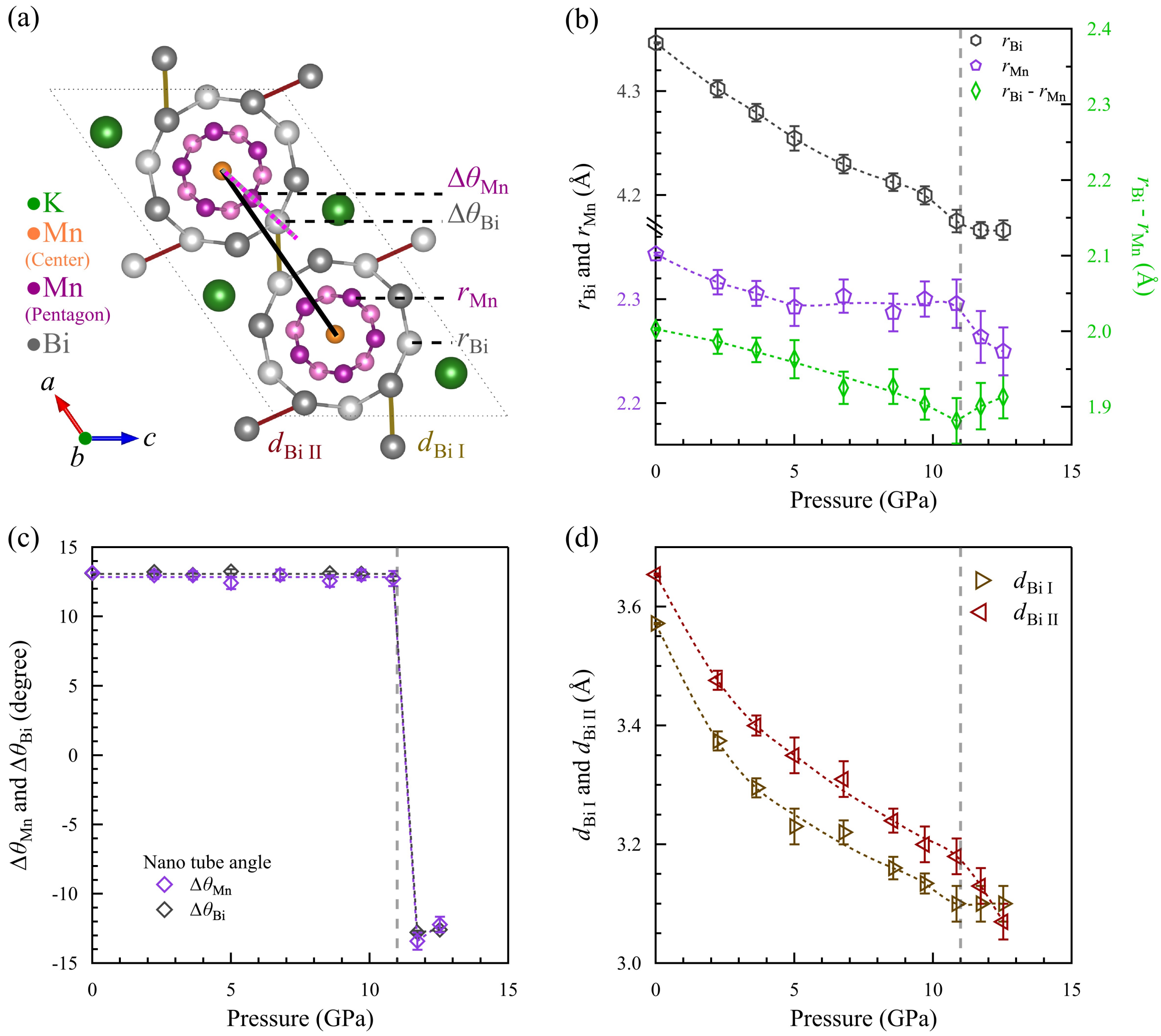

(a)
Δθ_Mn
Δθ_Bi
K
Mn
(Center)
Mn
(Pentagon)
Bi
r_Mn
r_Bi
a
b
c
d_Bi II
d_Bi I
(b)
r_Bi
r_Mn
r_Bi - r_Mn
r_Bi and r_Mn (Å)
r_Bi - r_Mn (Å)
Pressure (GPa)
(c)
Nano tube angle
Δθ_Mn
Δθ_Bi
Δθ_Mn and Δθ_Bi (degree)
Pressure (GPa)
(d)
d_Bi I
d_Bi II
d_Bi I and d_Bi II (Å)
Pressure (GPa)

# Figure 4

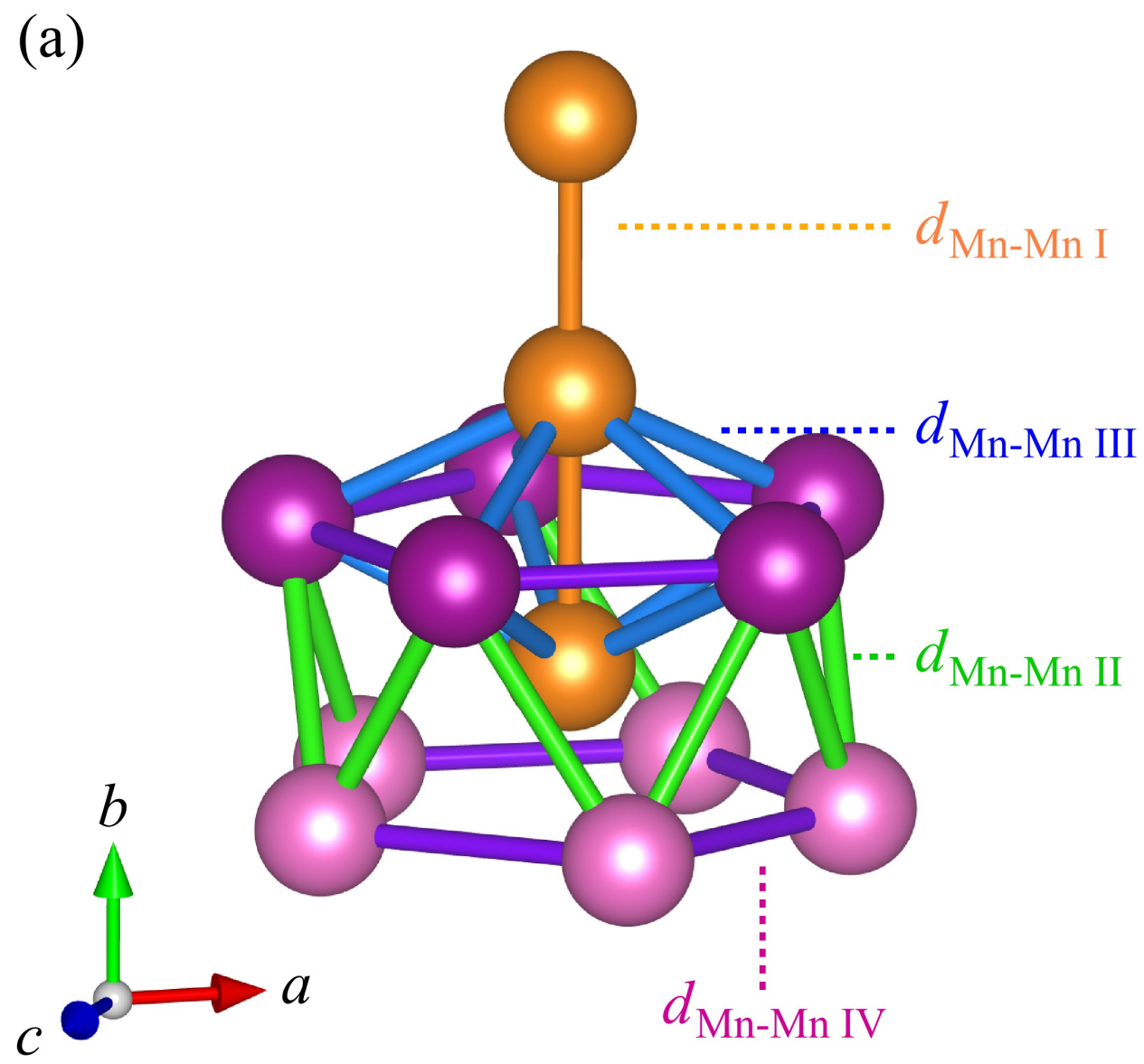
(a)
$d_{\text{Mn-Mn I}}$
$d_{\text{Mn-Mn III}}$
$d_{\text{Mn-Mn II}}$
$d_{\text{Mn-Mn IV}}$
b
a
c

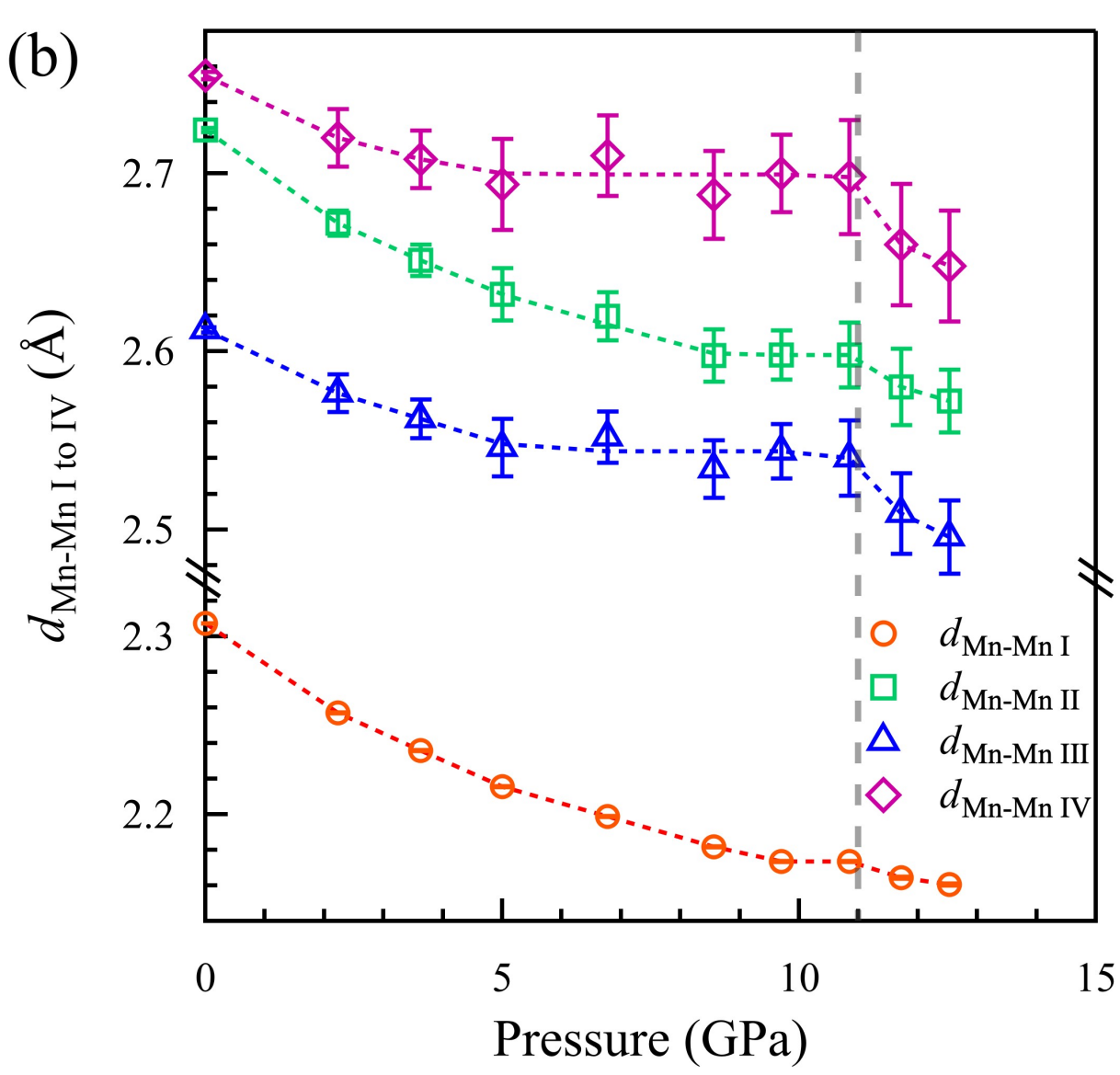
(b)
$d_{\text{Mn-Mn I to IV}}$ (Å)
2.7
2.6
2.5
2.3
2.2
$d_{\text{Mn-Mn I}}$
$d_{\text{Mn-Mn II}}$
$d_{\text{Mn-Mn III}}$
$d_{\text{Mn-Mn IV}}$
0
5
10
15
Pressure (GPa)

**Table 1 The lattice parameters, the fractional atomic positions, and equivalent isotropic displacement parameters $U_{eq}$ of $KMn_6Bi_5$ [a] under various pressures and room temperature.**

| Pressure (GPa) | | 0 | 2.23 | 3.62 | 5 | 6.77 | 8.56 | 9.7 | 10.85 | 11.72 | 12.53 |
|---|---|---|---|---|---|---|---|---|---|---|---|
| *a* (Å) | | 22.9934(10) | 22.26(4) | 22.00(4) | 21.73(6) | 21.54(4) | 21.27(3) | 21.17(4) | 21.11(5) | 20.89(5) | 20.88(5) |
| *b* (Å) | | 4.6147(1) | 4.5138(4) | 4.4714(5) | 4.4310(6) | 4.3978(4) | 4.3633(3) | 4.3469(4) | 4.3471(5) | 4.3288(8) | 4.3215(9) |
| *c* (Å) | | 13.3944(6) | 13.02(2) | 12.86(3) | 12.72(4) | 12.60(3) | 12.49(2) | 12.43(3) | 12.38(3) | 12.339(12) | 12.293(12) |
| $\beta$ (°) | | 124.591(6) | 124.9(3) | 124.9(3) | 124.9(4) | 125.0(3) | 125.0(3) | 125.1(3) | 125.8(4) | 125.8(2) | 125.7(3) |
| *V* ($Å^3$) | | 1170.01(1) | 1072(4) | 1038(5) | 1005(6) | 977(4) | 950(3) | 936(4) | 922(5) | 905(3) | 901(3) |
| K | *x* | 0.8687(2) | 0.8632(14) | 0.8612(15) | 0.8580(20) | 0.8580(20) | 0.8620(20) | 0.8630(20) | 0.8610(30) | 0.8580(50) | 0.8520(50) |
| | *z* | 0.1152(4) | 0.1100(30) | 0.1060(30) | 0.1010(40) | 0.1030(40) | 0.1110(40) | 0.1130(40) | 0.1110(60) | 0.1090(40) | 0.1060(30) |
| | $U_{eq}$ ($Å^2$) | 0.0342(9) | 0.031(3) | 0.028(3) | 0.021(3) | 0.31(4) | 0.022(3) | 0.030(4) | 0.027(5) | 0.044(11) | 0.041(9) |
| Mn 1 | *x* | 0.3624(13) | 0.3649(8) | 0.3660(9) | 0.3651(16) | 0.3681(14) | 0.3662(15) | 0.3714(13) | 0.3710(20) | 0.3330(30) | 0.3410(30) |
| | *z* | 0.6812(20) | 0.6838(16) | 0.6863(18) | 0.6840(30) | 0.6930(30) | 0.6880(30) | 0.6950(30) | 0.6950(40) | 0.7255(20) | 0.7250(15) |
| | $U_{eq}$ ($Å^2$) | 0.0142(4) | 0.0191(15) | 0.0160(15) | 0.026(3) | 0.025(2) | 0.017(2) | 0.017(2) | 0.020(3) | 0.026(6) | 0.014(4) |
| Mn 2 | *x* | 0.33411(13) | 0.3372(8) | 0.3363(8) | 0.3436(15) | 0.3364(12) | 0.3409(15) | 0.3389(13) | 0.3380(19) | 0.3680(30) | 0.3660(30) |
| | *z* | 0.4536(20) | 0.4549(16) | 0.4529(17) | 0.4630(30) | 0.4510(20) | 0.4560(30) | 0.4530(20) | 0.4520(40) | 0.5478(18) | 0.5472(17) |
| | $U_{eq}$ ($Å^2$) | 0.0142(4) | 0.0181(14) | 0.0126(14) | 0.020(2) | 0.018(2) | 0.021(2) | 0.016(2) | 0.018(3) | 0.019(5) | 0.020(5) |
| Mn 3 | *x* | 0.18774(12) | 0.1859(8) | 0.1843(8 | 0.1852(15) | 0.1813(13) | 0.1807(15) | 0.1802(13) | 0.1780(20) | 0.2410(30) | 0.2370(30) |
| | *z* | 0.2864(20) | 0.2835(16) | 0.2794(17) | 0.2810(30) | 0.2730(20) | 0.2700(30) | 0.2680(30) | 0.2650(40) | 0.3088(19) | 0.3084(17) |
| | $U_{eq}$ ($Å^2$) | 0.0128(4) | 0.0172(14) | 0.0130(14) | 0.022(2) | 0.021(2) | 0.020(2) | 0.016(2) | 0.022(3) | 0.025(6) | 0.018(4) |
| Mn 4 | *x* | 0.1293(12) | 0.1252(9) | 0.1243(10) | 0.1226(15) | 0.1205(14) | 0.1206(15) | 0.1202(13) | 0.1160(20) | 0.1210(40) | 0.1180(30) |
| | *z* | 0.4183(20) | 0.4144(17) | 0.4135(19) | 0.4120(30) | 0.4080(30) | 0.4100(30) | 0.4090(30) | 0.4040(40) | 0.3320(20) | 0.3289(15) |
| | $U_{eq}$ ($Å^2$) | 0.0129(4) | 0.0175(14) | 0.0164(16) | 0.017(2) | 0.025(2) | 0.015(2) | 0.015(2) | 0.021(3) | 0.034(7) | 0.014(4) |
| Mn 5 | *x* | 0.23726(12) | 0.2377(9) | 0.2365(10) | 0.2371(15) | 0.2393(16) | 0.2365(15) | 0.2396(13) | 0.2400(20) | 0.1780(30) | 0.1770(30) |
| | *z* | 0.6615(20) | 0.6641(18) | 0.6627(19) | 0.6650(30) | 0.6700(30) | 0.6660(30) | 0.6710(30) | 0.6710(40) | 0.5944(20) | 0.5903(17) |
| | $U_{eq}$ (Å$^2$) | 0.0139(4) | 0.0194(14) | 0.0169(15) | 0.019(2) | 0.029(3) | 0.021(2) | 0.014(2) | 0.017(3) | 0.024(5) | 0.018(4) |
| Bi 1 | *x* | 0.47596(3) | 0.4823(2) | 0.4850(2) | 0.4861(4) | 0.4882(4) | 0.4892(4) | 0.4896(3) | 0.4913(5) | 0.4920(8) | 0.4914(7) |
| | *z* | 0.65532(6) | 0.6596(4) | 0.6626(5) | 0.6633(7) | 0.6652(7) | 0.6658(7) | 0.6660(6) | 0.6684(9) | 0.8151(5) | 0.8157(4) |
| | $U_{eq}$ (Å$^2$) | 0.02055(17) | 0.0233(6) | 0.0196(6) | 0.0243(9) | 0.0264(9) | 0.0198(8) | 0.0166(6) | 0.0196(10) | 0.0270(18) | 0.0203(13) |
| Bi 2 | *x* | 0.2770(4) | 0.2767(2) | 0.2766(3) | 0.2765(4) | 0.2762(4) | 0.2775(4) | 0.2776(4) | 0.2774(5) | 0.3775(9) | 0.3778(8) |
| | *z* | 0.2063(6) | 0.2012(5) | 0.1995(5) | 0.1968(8) | 0.1950(7) | 0.1951(7) | 0.1952(7) | 0.1950(10) | 0.3322(5) | 0.3306(5) |
| | $U_{eq}$ (Å$^2$) | 0.02089(17) | 0.0233(6) | 0.0199(6) | 0.0242(9) | 0.0264(9) | 0.0200(8) | 0.0185(7) | 0.0206(11) | 0.0296(19) | 0.0246(14) |
| Bi 3 | *x* | 0.03987(3) | 0.0337(3) | 0.0319(3) | 0.0307(5) | 0.0299(4) | 0.0282(4) | 0.0272(4) | 0.0264(5) | 0.0924(8) | 0.0904(7) |
| | *z* | 0.16309(6) | 0.1578(5) | 0.1556(5) | 0.1546(9) | 0.1537(7) | 0.1511(7) | 0.1498(7) | 0.1496(10) | 0.0882(5) | 0.0869(4) |
| | $U_{eq}$ (Å$^2$) | 0.02233(17) | 0.0269(6) | 0.0234(6) | 0.0271(9) | 0.0288(9) | 0.0225(8) | 0.0200(7) | 0.0215(11) | 0.0270(18) | 0.0230(14) |
| Bi 4 | *x* | 0.09427(3) | 0.0924(2) | 0.0913(2) | 0.0907(4) | 0.0913(4) | 0.0906(4) | 0.0904(3) | 0.0906(5) | 0.0238(8) | 0.244(7) |
| | *z* | 0.58661(6) | 0.5904(4) | 0.5903(5) | 0.5913(8) | 0.5939(7) | 0.5943(7) | 0.5937(6) | 0.5936(10) | 0.4021(5) | 0.4027(4) |
| | $U_{eq}$ (Å$^2$) | 0.02076(17) | 0.0229(5) | 0.0193(6) | 0.0237(9) | 0.0261(9) | 0.0201(8) | 0.0176(7) | 0.0199(10) | 0.0285(19) | 0.0231(14) |
| Bi 5 | *x* | 0.36694(4) | 0.3716(2) | 0.3725(2) | 0.3730(4) | 0.3730(4) | 0.3741(4) | 0.3750(3) | 0.3751(5) | 0.2771(8) | 0.2770(8) |
| | *z* | 0.89374(6) | 0.9041(5) | 0.9069(5) | 0.9097(8) | 0.9114(7) | 0.9149(7) | 0.9175(7) | 0.9189(10) | 0.8595(5) | 0.8602(5) |
| | $U_{eq}$ (Å$^2$) | 0.02162(17) | 0.0251(6) | 0.0210(6) | 0.0254(9) | 0.0266(9) | 0.0220(8) | 0.0194(7) | 0.0222(11) | 0.0272(18) | 0.0238(14) |
| $R_1$ (%) | | 4.5 | 7 | 7.2 | 10.8 | 9.9 | 10.4 | 7.5 | 10.6 | 17.9 | 11.9 |

[a] Space group *C2/m*, K 4*i* (*x* 0 *z*), Mn1-Mn5 4*i* (*x* 0 *z*), Mn6 4*f* (1/4 1/4 1/2), Bi1-Bi5 4*i* (*x* 0 *z*).

**Table 2: The $r_{Mn}$, $r_{Bi}$, $\Delta\theta_{Mn}$, $\Delta\theta_{Bi}$, $d_{Mn\text{-}Mn\ I}$ to $d_{Mn\text{-}Mn\ IV}$, $d_{Bi\ I}$, and $d_{Bi\ II}$ under pressures.**

| Pressure (GPa) | $r_{Mn}$ (Å) | $r_{Bi}$ (Å) | $\Delta\theta_{Mn}$ (°) | $\Delta\theta_{Bi}$ (°) | $d_{Mn\text{-}Mn\ I}$ (Å) | $d_{Mn\text{-}Mn\ II}$ (Å) | $d_{Mn\text{-}Mn\ III}$ (Å) | $d_{Mn\text{-}Mn\ IV}$ (Å) | $d_{Bi\ I}$ (Å) | $d_{Bi\ II}$ (Å) |
|---|---|---|---|---|---|---|---|---|---|---|
| 0 | 2.344(1) | 4.346(1) | 13.17(3) | 13.121(5) | 2.30735(6) | 2.724(1) | 2.612(1) | 2.755(2) | 3.5716(13) | 3.6544(16) |
| 2.23 | 2.314(6) | 4.300(2) | 12.98(26) | 13.23(11) | 2.2569(3) | 2.672(7) | 2.577(10) | 2.720(16) | 3.374(16) | 3.476(16) |
| 3.62 | 2.304(6) | 4.279(2) | 12.98(29) | 13.01(12) | 2.2357(3) | 2.651(9) | 2.562(11) | 2.708(16) | 3.295(16) | 3.400(17) |
| 5 | 2.292(9) | 4.255(3) | 12.42(45) | 13.25(16) | 2.2155(4) | 2.632(15) | 2.546(16) | 2.694(25) | 3.23(3) | 3.35(3) |
| 6.77 | 2.305(8) | 4.230(2) | 13.0(4) | 13.0(4) | 2.1989(3) | 2.620(14) | 2.552(14) | 2.710(23) | 3.22(2) | 3.31(3) |
| 8.56 | 2.287(9) | 4.213(2) | 12.56(42) | 13.13(12) | 2.18165(16) | 2.598(15) | 2.534(16) | 2.688(25) | 3.160(19) | 3.24(2) |
| 9.7 | 2.297(8) | 4.200(2) | 13.0(4) | 13.09(12) | 2.1734(3) | 2.598(14) | 2.544(15) | 2.700(22) | 3.134(17) | 3.20(3) |
| 10.85 | 2.295(12) | 4.177(3) | 12.74(55) | 12.74(16) | 2.1735(3) | 2.598(18) | 2.540(21) | 2.698(32) | 3.10(3) | 3.18(3) |
| 11.72 | 2.263(12) | 4.164(3) | -13.42(61) | -12.77(12) | 2.1644(5) | 2.580(21) | 2.509(23) | 2.660(34) | 3.10(3) | 3.13(3) |
| 12.53 | 2.253(11) | 4.166(2) | -12.22(56) | -12.58(13) | 2.1607(5) | 2.572(18) | 2.496(21) | 2.648(31) | 3.10(3) | 3.07(3) |